# Accepted Manuscript

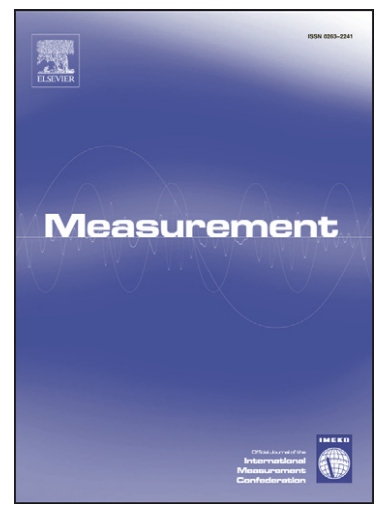

Alternative Procedures in Realizing of the High Frequency Power Standards
with Microcalorimeter and Thermoelectric Power Sensors

Emil Vremera, Luciano Brunetti, Luca Oberto, Marco Sellone

Please cite this article as: E. Vremera, L. Brunetti, L. Oberto, M. Sellone, Alternative Procedures in Realizing of
the High Frequency Power Standards with Microcalorimeter and Thermoelectric Power Sensors, *Measurement*
(2008), doi: 10.1016/j.measurement.2008.06.010

# Alternative Procedures in Realizing of the High Frequency Power Standards with Microcalorimeter and Thermoelectric Power Sensors

Emil Vremera, Luciano Brunetti, Luca Oberto, Marco Sellone

*Abstract*–The paper describes the realization process of the high frequency power standard based on a twin broadband microcalorimeter designed for effective efficiency measurement of coaxial power sensors with indirect heating thermocouples. The presented procedures involve microcalorimeter measurement and calibration steps, parameters computation and systematic errors correction for both long–term and accelerated algorithms. Also, a new method for determining the key parameters of high frequency power standards is proposed.

## 1. Introduction

The microcalorimeter is a measurement system conventionally considered as primary power standard. Basically, it is used for the effective efficiency measurements of the bolometer power sensors, like thermistor or barretter mounts, in order to obtain transfer standards [1]. The last microcalorimetric techniques improvements, both hardware and software, make from the thermocouple power sensors a good alternative to the bolometric power sensors. Moreover, the self–balancing direct bias current, which is required by the bolometer mounts, is now missing. Thus, the small relative variations of the measured temperature, detected from the heat developed by the power sensor, are avoided. The most critical component of this microcalorimeter is the feeding line, whose losses give the main measurement error. Its correction and the contribution to the system uncertainty can be determined in a microcalorimeter calibration step. This calibration is done by applying the concept of the perfect dummy load and by computing the feeding lines losses from $S$–parameters measurement [1], [2]. An important part of the system's uncertainty occurs from the $S$–parameters, obtained by vector network analyzer measurements, because of their great weights in the computation equations [2]. The maximum estimated uncertainty of the effective efficiency previously obtained as 2 %, [2], [3], [4], should be now four times lower in the same frequency band (between dc and 26.5 GHz).

In the design of the microcalorimeters, the twin type configuration is preferred because of its better thermal disturbance rejection. Moreover, various calibration methods are allowed by this structure. In Section 5 a new calibration method for this system is proposed.

## 2. Microcalorimeter and thermocouple power sensors

The measurement system we are considering is based on a dry adiabatic calorimeter, as the Figure 1 shows. The calorimetric thermal load consists of a pair of sensors, one of them – U – being the sensor under test (SUT). This is alternatively supplied with HF test power and LF or DC reference power through the insulating coaxial lines. To the other sensor – D –, playing the role of thermal dummy load (TDL), it is not applied any energy. It works only as a thermal reference mass, in order to obtain a differential configuration. This is efficiently rejecting the thermal fluctuations induced inside the microcalorimeter from the outside environment, even if the system has a high thermal insulation. As well, TDL could play the role of SUT without opening the microcalorimeter if a pair of real sensors is used. This possibility may ensure time saving in the calibration step. In Section 5, this idea is used in developing of a calibration process.

The thermal insulation of the microcalorimeter load is realized with a complex of aluminum–copper–aluminum shields separated by both polyurethane foam and air. The intermediate temperature of the copper–shield is actively controlled by means of Peltier heat pump that can stabilize the inner shield at ± 5 mK, as reported in [5]. Such thermostat is a more flexible alternative to that based on thick metal cylinder in thermostatic water bath, presented in [6], without having a significant depreciation of the temperature stability.

The feeding lines include two thin–wall coaxial segments fitted with connectors for an easy characterization by network analyzer measurements. These coaxial segments, named 1–st Section and 2–nd Section in Figure 1, assure high thermal insulation with the environment and good HF power propagation at the same time. From the feeding part communicating with the microcalorimeter environment and its associated instruments (1–st Section), only the inner conductor is designed in thin–wall technique. Such arrangement contributes to better filtering of the external thermal noise by conveying it to the thermal ground through the insulating line outer–wall. In order to minimize the mismatches, short insulating lines were preferred instead of the thermostated long and coiled lines [6].

The system has two thermometers, each of them being made with two independent Copper–Constantan thermopiles. One thermometer is adjusted for 3.5–mm coaxial SUTs, while the other is for 7–mm coaxial SUTs. Combining appropriately the signals received from thermopile pair, a measurement of the temperature difference between SUT and TDL is obtained. This is directly proportional to the SUT effective efficiency.

The system shown in Figure 1 can work with both bolometric and thermoelectric power sensors. At INRiM – Italy, the HF microcalorimeter is mostly used for thermoelectric power sensors [2]. All measurement steps are automated by computer control, excepting the connecting of the sensors to the feeding lines.

In the next sections are presented the mathematical models that allow determining the effective efficiency for the thermoelectric power sensors and the microcalorimeter calibration process. There are also highlighted the requirements to be fulfilled for maintaining the models validity.

### 3. Long–term measurement and calibration procedures

Before introducing the mathematical model of the system, we point out that after the substitution of the reference (REF ) power level – LF or DC – with an equivalent HF–power into the SUT, the thermopile output voltage $e$ begins to increase with an exponential trend, while $e$ decreases exponentially if HF–power is substituted with REF–power (LF or DC). The exponential increasing trend corresponds to the heating produced by the HF losses, while the exponential decreasing trend is relevant to a cooling owed to the loss removal we have with the power substitution HF – REF. However, $e$ tends always toward finite asymptotes. Only the $e$-values corresponding to a thermal equilibrium and therefore near the asymptotes are meaningful for our purpose. As these values are obtained after several time–constants $\tau$ of the system, they are called long–time measurements. For establishing a model that connects the thermometer response to the power sensor effective efficiency, it is convenient to divide the microcalorimeter thermopile voltage in two components [1], [2], [7], one related to the power sensor and the other to the feeding line:

$$e = \alpha \Delta T = \alpha R \left( k_1 P_S + k_2 P_{IL} \right), \tag{1}$$

where:

$\alpha$ – is the Seebek coefficient of thermopile junctions;

$R$ – a conversion constant depending on thermodynamic parameters of the thermal load;

$k_1, k_2$ – coefficients that describe the power separation between SUT and feeding line;

$P_S$, $P_{IL}$ – power dissipated in the sensor and in the insulating line respectively.

The thermopile response is $e_1$ when HF–power is supplied to the SUT mount and $e_2$ when an equivalent REF–power is substituted in it. Long-time measurements of $e_1$ and $e_2$, when the SUT output $U = const.$, are combined in the following ratio $e_R$

$$e_R = \frac{e_2}{e_1} = \left[ \left( \frac{P_S|_{REF}}{P_S|_{HF}} \right) \frac{1 + k \left( P_{IL} / P_S \right)|_{REF}}{1 + k \left( P_{IL} / P_S \right)|_{HF}} \right]_{U_{REF} = U_{HF}} = \frac{\eta_{eff}}{g}, \quad k = \frac{k_2}{k_1} \ . \tag{2}$$

This equation includes the power sensor *effective efficiency* $\eta_{eff}$ and the *microcalorimeter calibration factor g* respectively defined as:

$$\eta_{eff} = \left( \frac{P_S|_{REF}}{P_S|_{HF}} \right)\Bigg|_{U_{REF} = U_{HF}} \ , \tag{3}$$

$$g = \frac{1 + k a_{HF}}{1 + k a_{REF}}, \tag{4}$$

where the coefficients $a_{HF}$ and $a_{REF}$, related to the transmission parameters $S_{21}$ of the feeding lines, are determined by vectorial $S$–parameter measurements:

$$a_{HF\,or\,REF} = \frac{P_{IL}}{P_S}\Bigg|_{HF\,or\,REF} = \frac{1}{\left( 1 - |\Gamma_S|^2 \right)} \frac{1 - |S_{21}|^2}{|S_{21}|^2}\Bigg|_{HF\,or\,REF} \ . \tag{5}$$

The accuracy of the $g$ factor contributes almost completely to the total accuracy of the system, especially at the highest frequencies, where the insulating line losses are of the same order with the sensor losses. Therefore, an accurate microcalorimeter calibration is necessary. We may attempt to measure directly $g$ by reversing (2) if a load is available of well known effective efficiency, but we may also proceed as follows.

In our case, the microcalorimeter calibration requires both the transmission parameter $S_{21}$, and the $k$ ratio determination. The last one is a quite critical operation because it requires a change of the microcalorimeter configuration. The hot sensor and the dummy one are disconnected and substituted by two short circuits having the same thermal mass. This could change the thermal behaviour of the system, leading to an erroneous determination of the separation constants $k_1$ and $k_2$. Furthermore, the power entering the system must be maintained always constant, in order to obtain congruent measurements. However, supposing that the thermodynamic conditions of the system are maintained unchanged during this operation, the estimation of $k$ factor as a consequence of (1) and (2), is done by the equation:

$$k = \frac{e_{2_{SC}}}{e_2 - e_{2_{SC}}} \frac{P_S|_{REF;\,U_{REF}}}{P_{IL}|_{REF;\,SC}} \tag{6}$$

The index SC denotes the short circuit condition, at which the quantity $e_2$ have also to be measured.

The $k$–ratio is considered not depending of frequency and can be obtained by measurements at 1 kHz reference power. At this frequency it is easy enough to maintain constant the current and the losses on a feeding line correspond with those in the short–circuit termination case.

From (5) and (6), we compute the $g$ factor. Finally, the effective efficiency of the power sensor will be obtained from the equation (2).

The model just presented has an intrinsic problem in the determination of $S_{21}$. The feeding line extends from the SUT to the generator, well outside of the thermostat. It is not clear what section of this line really gives a contribution to the measurement process with its losses. Only the attenuation of line section that is inside of the thermostat body is considered effective. Though some doubts persist, because it is not easy to take into account the contribution of a coaxial transmission line inner conductor, this part is the main dissipative one at the REF power. A separation of losses in tree components, followed by a calibration based on a reference power sensor should have better results in terms of uncertainties [9].

For solving this problem, we have elaborated an alternative model, assuming the feeding line losses negligible at the REF frequency [2]. Having experimentally verified as true this assumption, we proceeded as following:

- the feeding line is terminated by SUT which is alternatively supplied with HF and REF powers of appropriated level for keeping the sensor output voltage, $U = const.$ In this case, it results:

$$e_1 = \alpha R \left( k_1 P_S + k_2 P_{IL} \right) \Big|_{HF; U_{HF}} \text{ and } e_2 = \alpha R k_1 P_S \Big|_{REF; U_{REF} = U_{HF}} ;$$

- the feeding line is short–circuited at the interface plane with the SUT connector and half of the previously HF power is injected into the system. It results: $e_{1_{SC}} = \alpha R k_2 P_{IL} \Big|_{HF; SC} .$

The effective efficiency of the thermoelectric power sensor, defined by (3), becomes

$$\eta_{eff} = \frac{e_2}{e_1 - e_{1_{SC}}} , \qquad (7)$$

where index SC is again relevant to the short circuit condition of the microcalorimeter test port. Similar results can be obtained by connecting the corresponding dummy load instead of SUT. The DC voltage $e_{1_{SC}}$ is clearly the correction term that accounts for the microcalorimeter losses. It would be appropriate as well to be able to maintain the thermodynamic equilibrium of the system and constant power losses on the feeding lines.

Equation (7) is important because it can show also the accuracy limit that a power standard can reach. For an ideal microcalorimeter, the correction term $e_{1_{SC}}$ becomes zero and therefore the accuracy of $\eta_{eff}$ is due to the measurement accuracy of two DC voltages; in our case, the thermopile output voltages $e_1$ and $e_2$ are of the mV order. Unfortunately, the calibration time becomes as long as the measurement time, the overall investigation time becoming two times greater, increasing also proportionally with the number of the requested test points.

## 4. Microcalorimeter calibration and accelerated measurements

After each power exchange (HF → REF or vice versa) the long–time measurement procedure implies a waiting state that should be about ten system time constants before sampling the microcalorimeter thermopile output [1]. With a time constant of 30 minutes, 8 test frequencies and requiring 10 independent results, the overall necessary time is over 800 hours. It is necessary more then a month, even with a continuous automatic measurement process. By reducing the measurement time to 1–3 time constants, the overall measurement time decreases considerably, but this method requires mathematical corrections [2], [7].

For relatively short switching time ($T_S$), the thermopile output does not provide directly the asymptotic values $e_1$ and $e_2$ requested by (1). These values may be obtained from the relative maximum and minimum values $e_M$ and $e_m$ of the thermopile output signal. The ratio $e_R$ of the thermopile output voltages, equivalent to the long–time measurement as defined in (2), is therefore given by:

$$e_R = \frac{e_m}{e_M} \cdot \frac{1-(e_M/e_m)\exp(-T_S/\tau)}{1-(e_m/e_M)\exp(-T_S/\tau)} = e_\tau H_\tau \ , \tag{8}$$

where $e_\tau$ is the ratio of the extreme values of the thermopile output voltage at the switching moments, and $H_\tau$ is a correction factor for the limited switching time.

The power separation coefficient ratio $k$ is obtained in the same manner as previously described. By using the same switching time as in the measurement case, this coefficient will result as:

$$k = \frac{e_{m_{SC}}}{e_m - e_{m_{SC}}} \cdot \frac{P_S\big|_{\text{REF; }U_{\text{REF}}}}{P_{\text{IL}}\big|_{\text{REF; SC}}} \tag{9}$$

From (4) and (9), we compute the $g$ factor. Finally, the effective efficiency of the power sensor is computed, in this case, by the following equation:

$$\eta_{eff} = e_R g = e_\tau H_\tau g \ . \tag{10}$$

Instead of using the correction factor $H_\tau$, it is possible to compute the thermopile output voltage ratio from the extrapolated values, which corresponds to the long–term measurements case. This yields the use of a nonlinear fitting algorithm as Levenberg–Marquard or Trust–region [10], [11], for a simplified model of the microcalorimeter. Gauss–Newton algorithm is not convergent for many data sets. If a more detailed model for the microcalorimeter is chosen, the fitting law becomes more complicated and new parameters will appear [11]. Fortunately, these parameters have a true physical signification and the computed corrected results become more accurate [10], [11], [12]. Since there are two main parts that contribute to the heat developing, SUT and the insulating line, a sum of two exponentials containing two different time–constants is suitable [11]:

$$fittedmodel\ (x) = a_1 \cdot \exp(-b_1 \cdot x) + a_2 \cdot \exp(-b_2 \cdot x) + c \ . \tag{11}$$

The Levenberg–Marquardt algorithm was used with the Least Absolute Residuals (LAR) regression, a common scheme to the Robust Least Squares method. The LAR scheme finds a curve that minimizes the absolute difference of the residuals, rather than the squared differences. Therefore, the extreme values have a smaller influence on the fit. The fitting process was realized in MATLAB software programme. As experimental values, a trust data set obtained at IEN (now INRiM) – Italy – in a key comparison at 20 GHz involving thermocouple power sensors was used [2], [13]. By using the fitting algorithm, a set of coefficients is obtained, as shown in Table 1. The large absolute values of $a$ coefficient are associated with the large time–constant $\tau_1$ and the small ones – with the small time–constant $\tau_2$. These time constants are 29 min. 4 s and 4 min. 28 s, respectively.

In Figure 2, the extrapolated values of the experimental accelerated measurement data, computed from the fitting laws, are represented together with the experimental data. A very good result of the fitting process is noticed; no visible differences are present. The value of the thermopile output voltages ratio $e_R$ is obtained now from $c$ coefficients ratio and $\eta_{eff}$ is computed with (2), (3) and (4) as in long–term measurement case.

In Table 1, the uncertainty was not associated with the fitting coefficients. However, from the MATLAB programme, every fitting coefficients set has statistic parameters associated, that describe the goodness of the fit. These parameters are:

– the sum of squares due to error ($SSE$): $SSE = \sum_{i=1}^{n} w_i (y_i - \hat{y}_i)^2$, where $w_i$ are the weights which determine how much each response value influences the final parameter estimates, $y_i$ is the observed value and $\hat{y}_i$ is the fitted response value;

– the $R$–square: $R-\text{square} = \dfrac{SSR}{SST} = 1 - \dfrac{SSE}{SST}$; $SSR$ and $SST$ are the sum of squares of the regression and the total sum of squares, respectively:

$$SSR = \sum_{i=1}^{n} w_i (\hat{y}_i - \bar{y})^2 \; , \quad SST = \sum_{i=1}^{n} w_i (y_i - \bar{y})^2 \; , \quad \text{where} \quad \bar{y} \text{ is the mean of the observed values;}$$

– the adjusted $R$–square: $\text{adjusted } R-\text{square} = 1 - \dfrac{SSE(n-1)}{SST(\nu)}$;

– the degrees of freedom for the adjusted $R$–square: the residual degrees of freedom are defined as the number of response values $n$ minus the number of fitted coefficients $m$ estimated from the response values: $\nu = n - m$, in this case $\nu = 90 - 5 = 85$;

– the root mean squared error ($RMSE$): $RMSE = s = \sqrt{MSE}$, where $MSE$ is the mean square error or the residual mean square, $MSE = \dfrac{SSE}{\nu}$.

In Table 2, the goodness parameters of the fit are shown. $RMSE$ of 0.22 % appears as upper limit for the investigated data set. It can be good enough, but it does not represent the uncertainty of the wanted quantity, the long–term thermovoltage ratio.

## 5. Measurement and calibration in a quasi–true twin–microcalorimeter working configuration

The twin microcalorimeter allows also another measurement procedure, by virtue of the intrinsic symmetries of its structure, at least in theory. In an ideal case, thermal contribution of the feeding paths, that is the main error source, becomes a common–mode quantity and can be rejected [14]. For being in such a case, the feeding paths must have equal insertion losses (same attenuation), the travelling energy on them must be maintained constant, the thermopiles must have the same thermoelectric response and the same power separation coefficients sets on both feeding lines. In a real case, we are more or less far from these conditions, [12], but a particular self-calibration method can be applied based on the measurement set–up schematized in Figure 3 if the thermal and electrical dissymmetry are made negligible with a refined design.

Two microwave coax switches, $S_1$ and $S_2$, appropriately connected to form a switching box, allow performing the measurement/calibration steps. The switching time of $S_1$ – between the reference power $P_{REF}$ and the test power $P_{HF}$ – is greater than the power sensor time constant. Conversely, the switching time of $S_2$ for power way selection – applied on SUT or TDL –must be short comparatively to the thermal time constant of the insulating lines $IL_1$, $IL_2$. The power level must be set two times greater than in the usual case in order to have the same total dissipated power and, consequently, the same thermopile maximum output voltage. If an adequate power–levelling algorithm, as that for keeping the same travelling energy on both paths, is applied, a part of the previous hypothesis is thus fulfilled. Measurement/calibration operations are repeated in two conditions:

– the $1^{st}$ step, when $IL_1$ is terminated by SUT and $IL_2$ is short–circuited by TDL;

– the $2^{nd}$ step, when $IL_1$ is short–circuited by TDL and $IL_2$ is terminated by SUT.

The original thermopile assembly, as Figure 1b shows, works now as an asymmetric two–channel thermometer. Considering the asymmetries between the thermopile sections as well as in the power separation coefficients $k$ and in the insulating lines losses, (3) will become now a more general equation related to the effective efficiency $\eta_{eff}$ of the SUT:

$$1/\eta_{eff} = \frac{P_{HF}}{P_{REF}}\bigg|_{U_{REF}=U_{HF}} = \frac{e_\theta(P_{HF})}{e_\theta(P_{REF})} - \frac{\alpha_1 R_1 k_{21}\left(P_{IL_{11}} - P_{IL_{12}}\right) + \alpha_2 R_2 k_{22}\left(P_{IL_{21}} - P_{IL_{22}}\right)}{e_\theta(P_{REF})}, \tag{12}$$

with $e_\theta(P) = \left(\alpha_1 R_1 k_{11} + \alpha_2 R_2 k_{12}\right)P$. The new introduced index is referring to the feeding path, while the second index digit of $P_{IL}$ corresponds to the measurement step. The last ratio in (12) can be minimized by a suitable design of the microcalorimeter, confirmed succeeding by the $S$–parameters measurement and an excellent power levelling process. By controlling all implied quantities, this term becomes really insignificant. Thus, the effective efficiency equation looks like the ideally simplified equation:

$$\eta_{eff} = \frac{P_{REF}}{P_{HF}}\bigg|_{U_{REF}=U_{HF}} = \frac{e_\theta(P_{REF})}{e_\theta(P_{HF})}. \tag{13}$$

This expression of the effective efficiency is attractive because no correction term is present. In the uncertainty budget there will be only two terms, but every quantity in this ratio is obtained from two data sets corresponding to two measurement steps. Moreover, the neglected term can be estimated by its upper limit and will be assumed as type B uncertainty described by the rectangular distribution, [15]. A special attention must be paid to the thermopile, in order to be optimized from many points of view: signal/noise ratio and signal/offset ratio, absorbed energy and measuring sequences, resolution and accuracy, [12], [16].

## 6. Practical results and discussion

Several accelerated measurements were performed recently at INRiM on a previously calibrated thermocouple power sensor NRV–Z52, SN: 828172/011 (Rohde & Schwarz). The experiment was focused to check the described methods. A single power level of 3 mW and three frequencies in the operating band were chosen: 1, 10 and 26 GHz. Two values were used for the switching period $T_S$, 30 and 90 minutes, corresponding to about one, respectively three time–constant. The measurements were repeated for the short–circuited feeding line case, corresponding to the microcalorimeter calibration. Four REF – HF completely switching cycles were performed for every case, meaning an investigation time of 2 hours in the high accelerated measurements and 6 hours in the other case. Two modalities were used in $\eta_{eff}$ computation:

– the values of the thermopile output voltages used in (7) were corrected as in $e_R$ ratio case, (8);

– by fitting the numerator and the denominator from (7), using (11), therefore obtaining the equivalent to the necessary long-term values.

Table 3 highlights the effective efficiency values in two situations: true–considered and computed ones. The computed values have their associated relative uncertainties, rel $u$, obtained as following of a summary uncertainty evaluation process on every seven data series corresponding to every case.

In this table, the optimum considered results were marked; these are close by the wanted quantity and are obtained from the best repeatability data series. As a rule, better results correspond to a switching time of 90 minutes. Moreover, by using a fitting algorithm for obtaining the long–term values, the results become closer to the true–considered ones and their repeatability is improved. The other used switching period leads to results not too far of the true–considered ones, but the repeatability is visibly compromised and it cannot be improved through the fitting process. For reaching such a target, it must measure more accurately the thermopile output voltage, with a good repeatability, and in very good electromagnetic compatibility conditions.

## 7. Conclusion

All presented methods can be efficiently applied in realization of the high frequency power standard using a twin broadband microcalorimeter designed for coaxial thermocouple power sensors. The accelerated measurement methods lead to good results with insignificant increase of the overall uncertainty with respect to classical long–term measurements. The best procedure is combining the accelerated measurements with data fitting algorithms, though this requires a lot of assisted computations.

We highlight a process that allows obtaining the calibration data sets simultaneously both for microcalorimeter and for transfer power standard. This method does not require a data correction step and seems to be a real alternative to the methods based on corrections, although it is quite difficult from the hardware point of view. Despite high requirements, the method is under test at INRiM–Italy by using thermoelectric power sensors independently calibrated. In terms of measuring time, the quasi–true twin–microcalorimeter is not visible competitive if compared against the other accelerated measurement methods. However, an improved uncertainty is expected due to the simplicity of the equation (13). In addition, the connector repeatability may be now considered as included in the uncertainty budget due to the mount–dismount operations performed in the feeding paths permutation step.

Table 1. Fitting coefficients series for a sum of two exponentials as fitting function; experimental data set was delivered in a key comparison, [2], [13].

| Fitt. set→ / Fitt. coeff.↓ | 1 | 2 | 3 | 4 | 5 | 6 | 7 | 8 | 9 | 10 | 11 | 12 | Av. set 1 | Av. set 2 | Obs. |
|---|---|---|---|---|---|---|---|---|---|---|---|---|---|---|---|
| $a_1$ [$10^{-5}$V] | -0.6734 | -0.0730 | -0.6365 | -0.0584 | -0.6428 | -0.0674 | -0.6277 | -0.0585 | -0.6332 | -0.0733 | -0.6242 | -0.0673 | -0.6396 | -0.0663 | |
| $a_2$ [$10^{-5}$V] | 0.0806 | 0.6459 | 0.0618 | 0.6344 | 0.0649 | 0.6366 | 0.0515 | 0.6351 | 0.0643 | 0.6435 | 0.0508 | 0.6434 | 0.0623 | 0.6398 | |
| $b_1$ [1/min.] | 0.0342 | 0.1874 | 0.0342 | 0.2047 | 0.0345 | 0.1863 | 0.0337 | 0.2220 | 0.0345 | 0.1929 | 0.0337 | 0.1914 | 0.0344 | | $\tau_1$ = 29 min. 4 s |
| $b_2$ [1/min.] | 0.3596 | 0.0356 | 0.2071 | 0.0343 | 0.2154 | 0.0352 | 0.2262 | 0.0340 | 0.2245 | 0.0349 | 0.2694 | 0.0344 | 0.2239 | | $\tau_2$ = 4 min. 28 s |
| $c$ [$10^{-5}$V] | 2.4390 | 1.8420 | 2.4400 | 1.8400 | 2.4450 | 1.8540 | 2.4500 | 1.8460 | 2.4430 | 1.8450 | 2.4440 | 1.8420 | 2.4435 | 1.8448 | $e_R$ = **0.7550** |
| $e_R$ | | 0.7552 | 0.7549 | 0.7541 | 0.7526 | 0.7583 | 0.7567 | 0.7535 | 0.7556 | 0.7552 | 0.7549 | 0.7537 | **0.7550** | | rel u($e_R$) = 0.066% |

Table 2. Goodness parameters associated with the fitting coefficients sets.

| Fitting set → / Good. param.↓ | 1 | 2 | 3 | 4 | 5 | 6 | 7 | 8 | 9 | 10 | 11 | 12 |
|---|---|---|---|---|---|---|---|---|---|---|---|---|
| *SSE* | 0.00042 | 0.00008 | 0.00011 | 0.00009 | 0.00011 | 0.00006 | 0.00009 | 0.00009 | 0.00009 | 0.00011 | 0.00014 | 0.00013 |
| *R*-square | 0.99983 | 0.99996 | 0.99995 | 0.99996 | 0.99995 | 0.99997 | 0.99996 | 0.99996 | 0.99996 | 0.99995 | 0.99993 | 0.99994 |
| Adj. *R*-square | 0.99982 | 0.99996 | 0.99995 | 0.99996 | 0.99995 | 0.99997 | 0.99995 | 0.99996 | 0.99996 | 0.99995 | 0.99993 | 0.99994 |
| *RMSE* | 0.0022 | 0.0010 | 0.0011 | 0.0010 | 0.0011 | 0.0008 | 0.0010 | 0.0010 | 0.0010 | 0.0011 | 0.0013 | 0.0012 |

Table 3. Comparison between the true-considered effective efficiency and the computed ones based on accelerated measurement methods, Section 4.

| f [GHz] | $\eta_{eff}$ True | $\eta_{eff}$ Comp. | rel u [%] | $\Delta(\eta_{eff})$ | Obs. |
|---|---|---|---|---|---|
| 1 | **0.9815** | 0.9833 | 0.0171 | 0.0018 | $H_t$, $T_S$=90min |
| | | 0.9796 | 0.3548 | -0.0019 | $H_t$, $T_S$=30min |
| | | 0.9844 | 0.0133 | 0.0029 | Fitt., $T_S$=90min |
| 10 | **0.9286** | 0.9280 | 0.0187 | -0.0006 | $H_t$, $T_S$=90min |
| | | 0.9333 | 0.0621 | 0.0047 | $H_t$, $T_S$=30min |
| | | 0.9287 | 0.0068 | 0.0001 | Fitt., $T_S$=90min |
| | | 0.9320 | 0.1138 | 0.0034 | Fitt., $T_S$=30min |
| 26 | **0.8567** | 0.8538 | 0.1569 | -0.0029 | $H_t$, $T_S$=90min |
| | | 0.8546 | 0.3278 | -0.0021 | $H_t$, $T_S$=30min |
| | | 0.8545 | 0.1688 | -0.0022 | Fitt., $T_S$=90min |

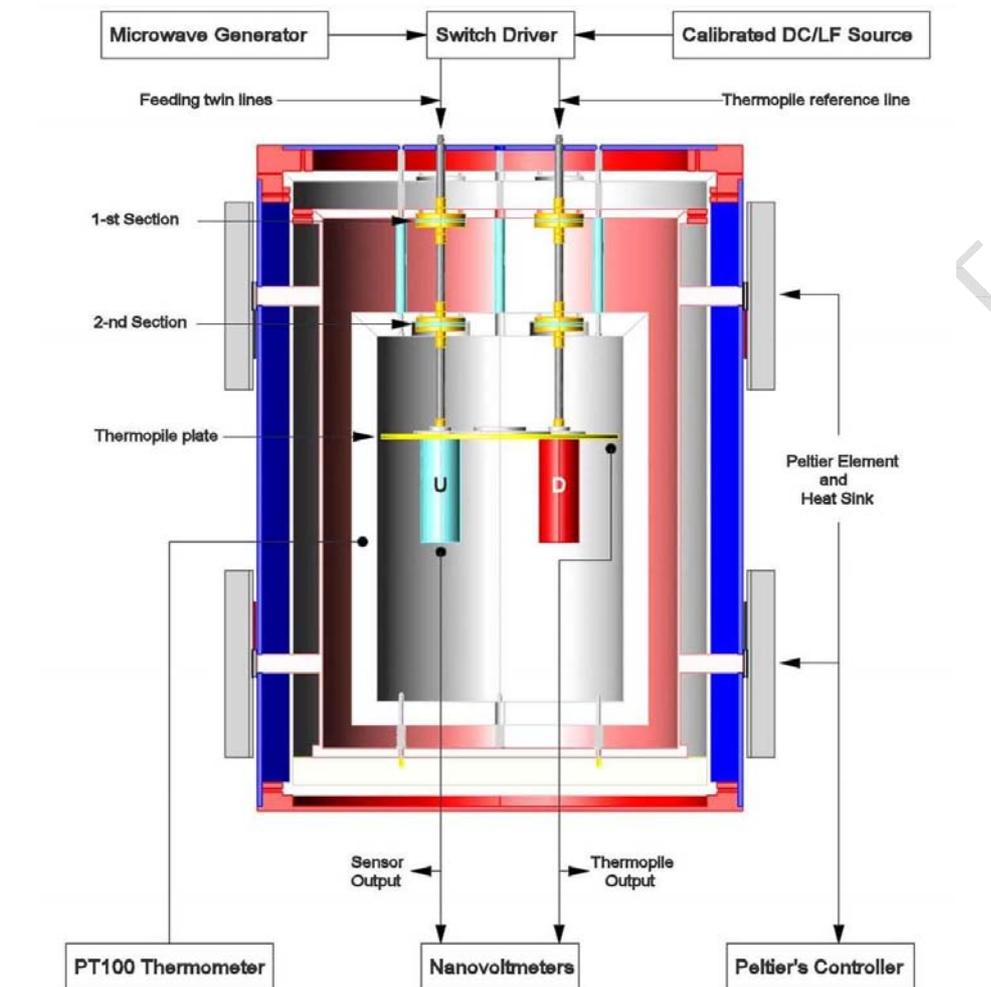

a)

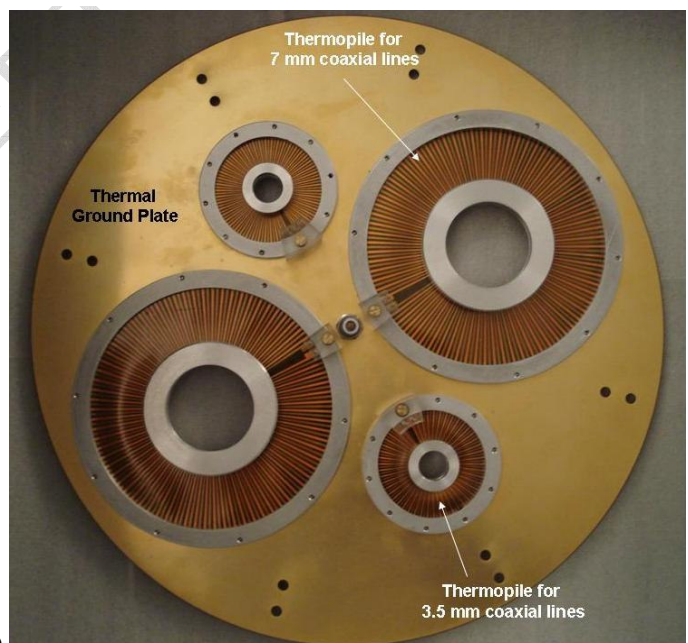

b)

Figure 1: a. Scaled cross section of the INRiM coaxial microcalorimeter;
b. Thermopile assemblies for twin coaxial microcalorimeter.

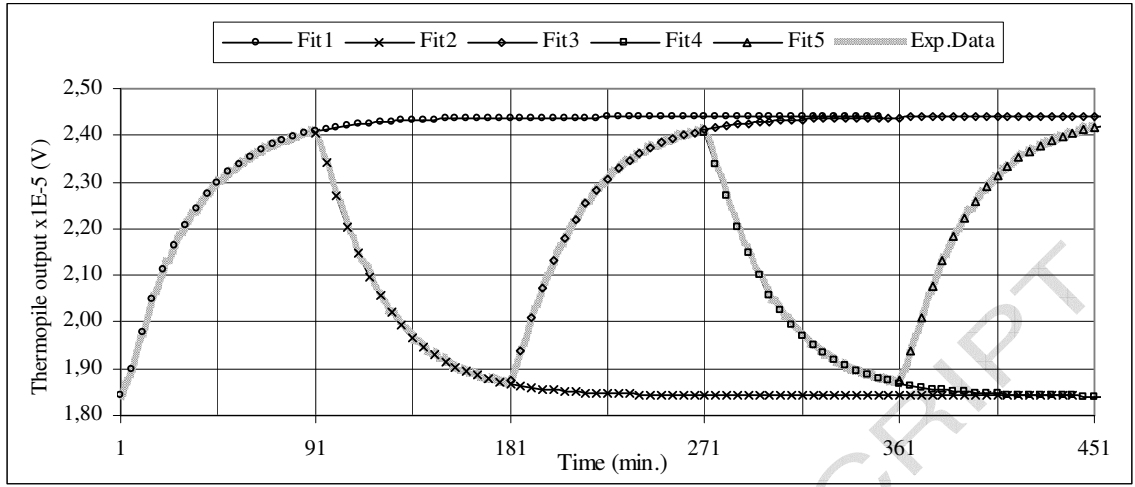

Figure 2. Extrapolated thermopile output waveform resulted by nonlinear fitting law method applied in the accelerated-measurements case for time constants $\tau_1 = 29$ min. 4 s and $\tau_2 = 4$ min. 28 s, a switching time $T_S = 90$ minutes and an extrapolation time interval of 305 minutes.

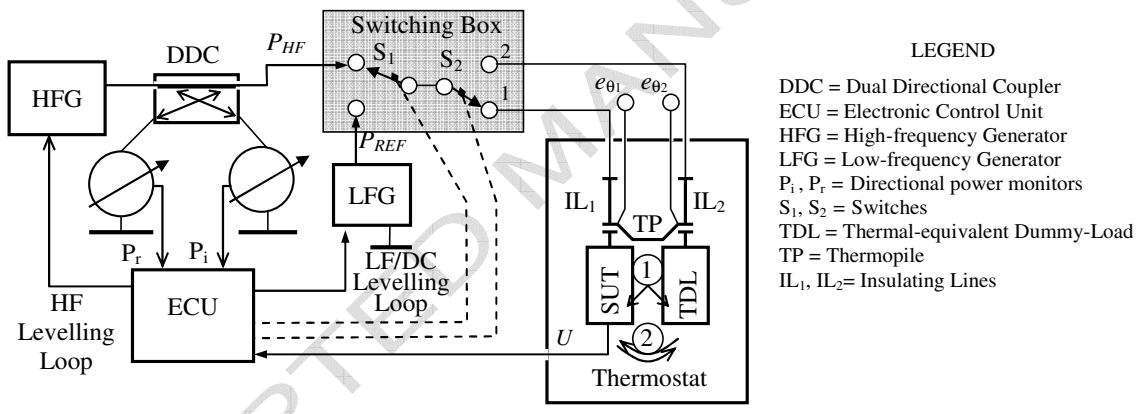

Figure 3. Quasi-true twin-microcalorimeter working configuration: the instrumentation set-up